\documentclass[a4paper,11pt]{article}

\usepackage[dvips]{graphicx,psfrag}
\usepackage{amssymb}
\usepackage{amsmath}

\makeatletter

 \@addtoreset{equation}{section}
\makeatother
\newcommand{\beq}{\begin{eqnarray}}
\newcommand{\eeq}{\end{eqnarray}}

\begin{document}

\begin{titlepage}

\vspace*{-1cm}
\begin{flushright}
OU-HET 527\\
KANAZAWA-05-06\\
{\tt hep-th/0505056}\\
May 2005
\end{flushright}

\begin{center}
{\LARGE\bf
Large-N Analysis of Three Dimensional Nonlinear Sigma Models
}

\setcounter{footnote}{0}

{\renewcommand{\thefootnote}{\fnsymbol{footnote}}
{\large\bf Kiyoshi Higashijima $^1$\footnote{
     E-mail: {\tt higashij@phys.sci.osaka-u.ac.jp}},
 Etsuko Itou$^2$\footnote{
     E-mail: {\tt itou@hep.s.kanazawa-u.ac.jp}} and 
 Makoto Tsuzuki$^3$\footnote{
     E-mail: {\tt makoto.tsuzuki@toshiba.co.jp}}

}}


{\sl
$^1$Department of Physics,
Graduate School of Science, Osaka University,\\ 
Toyonaka, Osaka 560-0043, Japan \\
$^2$Institute of Theoretical Physics Kanazawa University\\
Kakumamachi Kanazawa 920-1192, Japan \\
$^3$Toshiba Corporation Semiconductor Company\\
Kawasaki,Kanagawa 212-8583,Japan
}

\end{center}

\begin{abstract}
Non-perturbative renormalization group approach suggests that a large class of nonlinear sigma models are renormalizable in three dimensional space-time, while they are non-renormalizable in perturbation theory. ${\cal N}=2$ supersymmetric nonlinear sigma models whose target spaces are Einstein-K\"{a}hler manifolds with positive scalar curvature belongs to this class. hermitian symmetric spaces, being homogeneous, are specially simple examples of these manifolds.
To find an independent evidence of the nonperturbative renormalizability of these models, the large N method, another nonperturbative method, is applied to $3$-dimensional ${\cal N}=2$ supersymmetric nonlinear sigma models on the target spaces $CP^{N-1}=SU(N)/[SU(N-1)\times U(1)]$ and $Q^{N-2}=SO(N)/[SO(N-2)\times SO(2)]$, two typical examples of hermitian symmetric spaces.

We find that $\beta$ functions in these models agree with the results of the nonperturbative renormalization group approach in the next-to-leading order of $1/N$ expansion, and have non-trivial UV fixed points.
The $\beta$ function of the $Q^{N-2}$ model receives a nonzero correction in the next-to-leading order of the $1/N$ expansion.

We also investigate the phase structures of our models.
The $CP^{N-1}$ model has two phases; $SU(N)$ symmetric and asymmetric phase.
The $Q^{N-2}$ model has three phases; Chern-Simons, Higgs and $SO(N)$ broken phases.
In the Chern-Simons and Higgs phase, $SO(N)$ symmetry remains unbroken and all dynamical fields becomes massive.  An auxiliary gauge field also acquires mass, through an induced Chern-Simons term in the Chern-Simons phase, and through the vacuum expectation value of a di-quark bound state in the Higgs phase.

\end{abstract}

\end{titlepage}

\pagestyle{plain}

\section{Introduction}
Non-perturbative analyses reveal non-trivial features of quantum field theories which cannot be found within the perturbative analysis. One of the powerful non-perturbative method is the Wilsonian renormalization group (WRG) approach \cite{Wilson Kogut,Wegner and Houghton,Aoki,HI}. In \cite{HI-3dim}, WRG method was applied to three dimensional ${\cal N}=2$ supersymmetric nonlinear sigma models (NL$\sigma$M). Three dimensional NL$\sigma$Ms are perturbatively non-renormalizable according to the power counting. In the WRG approach, the renormalizability of NL$\sigma$Ms is equivalent to the existence of a nontrivial continuum limit, $\Lambda \rightarrow \infty$. When the ultraviolet (UV) cutoff $\Lambda$ tends to infinity, we have to fine tune the coupling constant to the critical value at the UV fixed point, so as to keep the observable quantities finite. Therefore, it is important to show the existence of the UV fixed point without using the perturbation theory. \footnote{${\cal N}=2$ supersymmetry in three dimensions requires the target manifolds of NL$\sigma$Ms are K\"{a}hler manifolds.} 
One of the main results in the paper \cite{HI-3dim} is the existence of non-trivial ultraviolet (UV) fixed point in the NL$\sigma$Ms whose target space are Einstein-K\"ahler manifolds with positive scalar curvature. This result implies that these models are renormalizable in the WRG approach.
In the WRG method, the renormalization group equation is an exact equation for the most general effective action which includes an infinite number of coupling constants.  To make this equation more tractable, we usually expand the effective action in powers of derivatives and retain the first few terms \cite{Morris}. In the WRG analysis in \cite{HI-3dim}, the most general effective action is approximated by truncating at the second order of derivative expansion.
Since it is difficult to justify the validity of this approximation directly, we apply another powerful non-perturbative method, the large $N$ method, to three dimensional ${\cal N}=2$ supersymmetric nonlinear sigma models (NL$\sigma$M) to show the existence of UV fixed point. 

The $1/N$ expansion method has been applied to show the nonperturbative renormalizability of the three dimensional Gross-Neveu model \cite{Gross-Neveu, Parisi, Shizuya, Rosenstein-War}. and NL$\sigma$M with $O(N)$ symmetry \cite{Arefeva, Rosenstein}. In NL$\sigma$M, however, quadratic terms of an auxiliary field, introduced as a lagrange multiplier field, appears as counter terms and destroy the renormalizability in the strict sense.
Three dimensional $O(N)$ symmetric NL$\sigma$M with ${\cal N}=1$ supersymmetry has been shown to be renormalizable in the large $N$ method because of the supersymmetry \cite{Koures}. The $\beta$ function of this model is known to have a nontrivial UV fixed point. NL$\sigma$M on the complex projective manifold ${\bf C}P^{N-1}$ has been formulated by introducing a auxiliary gauge field \cite{Witten, DAdda, large-N}. If we impose the gauge invariance to eliminate the possible digergent graphs, ${\bf C}P^{N-1}$ model is renormalizable in the large N method.
Inami, Saito and Yamamoto have explicitly shown that ${\cal N}=2$ supersymmetricNL$\sigma$M on ${\bf C}P^{N-1}$ has no next-to-leading order contribution to the $\beta$ function in the $1/N$ expansion \cite{Inami}, confirming the previous argument \cite{Morozov, Gracey}.

In the large N method, we introduce N copies of dynamical fields and appropriate auxiliary fields to linearize the lagrangian of the dynamical fields, so that these dynamical fields can be integrated out in the path integral formalism \cite{Moshe}. Auxiliary field formulation of supersymmetric NL$\sigma$Ms has been found for a special class of K\"{a}hler-Einstein manifolds called hermitian symmetric spaces \cite{large-N}. Hermitian symmetric spaces are homogeneous coset spaces with K\"{a}hler structure. Among the hermitian symmetric spaces, ${\bf C}P^{N-1}=SU(N)/[SU(N-1)\times U(1)]$ and $Q^{N-2}=SO(N)/[SO(N-2)\times SO(2)]$ are suitable for large N analysis. 
The $\beta$ function for $CP^{N-1}$ model agrees with the result of the WRG analysis. However, the WRG result suggests that the $\beta$ function for the other ${\cal N}=2$ supersymmetric NL$\sigma$Ms have next-to-leading order corrections in the $1/N$ expansion.

In this paper, we investigate the $Q^{N-2}$ model, which is formulated explicitly with auxiliary fields in paper \cite{HKNT}.
The $\beta$ function of $Q^{N-2}$ model obtained by the WRG analysis suggests that there is a next-to-leading order correction for the $\beta$ function in $1/N$ expansion. Therefore, we explicitly calculate the $\beta$ function of the $Q^{N-2}$ model.
Furthermore we study the phase structure of the $CP^{N-1}$ and the $Q^{N-2}$ models.
The ${\bf C}P^{N-1}$ model is known to have two phases; $SU(N)$ symmetric and asymmetric phase.
In the symmetric phase, all dynamical fields acquire a mass, while they play the role of the massless Nambu-Goldstone bosons and their superpartners in the broken phase. The $Q^{N-2}$ model turns out to have three phases; Chern-Simons, Higgs and $SO(N)$ broken phases.
In the Chern-Simons and Higgs phase, $SO(N)$ symmetry remains unbroken and all dynamical fields becomes massive, while there are $(N-2)$ massless Nambu-Goldstone bosons and their superpartners in the $SO(N)$ broken phase.  In the Chern-Simons phase, a auxiliary gauge fields becomes massive through an induced Chern-Simons term, while in Higgs phase the gauge field acquire the mass by the Higgs mechanism through the vacuum expectation value of a di-quark bound state.

This paper is organized as follows.
First, we will reconfirm the $\beta$ function of ${\bf C}P^{N-1}$ model has no next-to-leading contributions in $1/N$ expansion, and will compare the $\beta$ function with the WRG result in \S.\ref{sec-CPN}.
We will also investigate the phase structures of ${\bf C}P^{N-1}$ in this section.
Next we will investigate the phase structure and the $\beta$ function of the $Q^{N-2}$ model in \S.\ref{sec-QN}.

\section{The ${\bf C}P^{N-1}$ model} \label{sec-CPN}
In this section, we review the large-$N$ analysis of the ${\bf C}P^{N-1}$ model.
The $\beta$ function of this model has no next-to-leading order correction because of ${\cal N}=2$ supersymmetry \cite{Inami}.
\subsection{The auxiliary field formulation of ${\bf C}P^{N-1}$ model}
Let us introduce chiral superfields $\Phi^i(x,\theta)\ (i=1,2, \cdots, N)$ belonging to a fundamental representation of $G=SU(N)$, the isometry group of ${\bf C}P^{N-1}$. 
We also introduce $U(1)$ gauge symmetry 
\begin{equation}
 \Phi(x,\theta) \longrightarrow \Phi'(x,\theta)
  = e^{i \Lambda(x,\theta)} \Phi(x,\theta) \label{eqn:projective}
\end{equation}
to require that $\Phi(x,\theta)$ and $\Phi'(x,\theta)$ are 
physically indistinguishable. 
With a complex chiral superfield, 
$e^{i \Lambda(x,\theta)}$ is an arbitrary complex number. 
$U(1)$ gauge symmetry is thus complexified to $U(1)^{\bf C}$. 
The identification 
$\Phi \sim \Phi'$ defines the complex projective space ${\bf C}P^{N-1}$.
In order to impose local $U(1)$ gauge symmetry, 
we have to introduce a $U(1)$ gauge field 
$V(x, \theta, \bar{\theta})$ with 
the transformation property 
$e^{-V} \longrightarrow e^{-V} e^{-i\Lambda + i\Lambda^*}$.
$V(\theta,\bar{\theta},x)$ is a real scalar superfield\cite{WB}, and defined by dimensional reduction from $4$-dimensional ${\cal N}=1$ to $3$-dimensions. 
Then the lagrangian with a local $U(1)$ gauge symmetry is given by
\beq
{\cal L}=\int d^4 \theta (\Phi^i \Phi^{\dag \bar{i}} e^{-V} +c V), \label{CPN-kahler2}
\eeq
where the last term $V$ is called the Fayet-Illiopoulos D-term.
In this model, the gauge field $V(x,\theta,\bar{\theta})$ 
is an auxiliary superfield without kinetic term. 
The K\"ahler potential $K(\Phi, \Phi^*)$ 
is obtained by eliminating $V$ 
using the equation of motion for $V$
\begin{equation}
  {\cal L} = \int d^2\theta d^2\bar{\theta} K(\Phi,\Phi^*) 
  = c\int d^2\theta d^2\bar{\theta} 
     \log{({\vec{\Phi}}^*\cdot\vec{\Phi})}.
  \label{eqn:kahlerpot}
\end{equation}
This K\"ahler potential reduces to the standard Fubini-Study metric of 
${\bf C}P^{N-1}$ 
\begin{equation}
  K(\phi,\phi^*)=c\log{(1+\sum_{i=1}^{N-1}\Phi^{i*}\Phi^i)}
   \label{eqn:fsmetric}
\end{equation}
by a choice of gauge fixing
\begin{equation}
  \Phi^N(x,\theta)=1.\label{eqn:cpgauge}
\end{equation}
The global symmetry $G=SU(N)$, the isometry of the target space
${\bf C}P^{N-1}$, is linearly realized on our $\Phi^i$ fields and our 
lagrangian (\ref{CPN-kahler2}) with auxiliary field $V$ is manifestly 
invariant under $G$. The gauge fixing condition (\ref{eqn:cpgauge}) is
not invariant under $G=SU(N)$ and we have to perform an appropriate 
gauge transformation simultaneously to compensate the change of
$\Phi^N$ caused by the $SU(N)$ transformation. Therefore the global 
symmetry $G=SU(N)$ is nonlinearly realized in the gauge fixed theory. 
In this sense, our lagrangian (\ref{CPN-kahler2}) use the linear 
realization of $G$ in contrast to the nonlinear lagrangian (\ref{eqn:fsmetric}) in terms 
of the K\"ahler potential which use the nonlinear realization of $G$.

Instead of fixing the gauge to eliminate one component of chiral superfield, we can eliminate the chiral and and anti-chiral components in the gauge superfield $V(x,\theta,\bar{\theta})$. In our discussion below, we employ this Wess-Zumino gauge. Then the gauge superfield is written by component fields:
\beq
V=\bar{\theta} \gamma^\mu \theta v_{\mu} +\bar{\theta} \theta M +\frac{1}{2}\theta^2\bar{\theta}\lambda +\frac{1}{2} \bar{\theta}^2\bar{\lambda}\theta  + \frac{1}{4}\theta^2\bar{\theta}^2D, \nonumber
\eeq
where $v_{\mu}$ is a gauge field of three dimensional theory and the scalar field $M$ corresponds to the fourth component of $4$-dimensional vector field. Note that the real part of the scalar component of the gauge transformation $\Lambda(x,\theta)$ is not fixed yet with this gauge choice. We have to fix this residual gauge freedom to calculate the $1/N$ correction where this auxiliary gauge field begins to propagate. 
In order to make the Lagrangian of order $N$ in the large $N$ limit, we take the coefficient Fayet-Illiopoulous D-term $c=N/g^2$ and keep $g^2$ fixed when we take the limit of $N \rightarrow \infty$.
With this gauge choice, the global $SU(N)$ symmetry coming from the isometry of the target manifold is linearly realized, while the supersymmetry is realized non-linearly.
Furthermore, the action is invariant under $U(1)$ gauged symmetry generated by the real part of the gauge function $\Lambda(x,\theta)$, with the assignment of the $U(1)$ charge
\beq
[\Phi]=1 , \hspace{0.5cm} [\Phi^\dag]=-1.\label{U(1)sym}
\eeq

Integrating out the Grassmann coordinates $\theta$ and $\bar{\theta}$, the action (\ref{CPN-kahler2}) is written by using the component fields:
\beq
{\cal L}&=&\partial_\mu \varphi^{*i } \partial^\mu \varphi^i +i \bar{\psi}^i \partial\llap / \psi^i +F^i F^{i*} -[i (\varphi^{*i} \partial_\mu \varphi^i -\varphi^i \partial_\mu \varphi^{i*}) +\bar{\psi}^i \gamma_\mu \psi^i ]v^\mu +v^\mu v_\mu \varphi^{*i } \varphi^i \nonumber\\
&& - M^2 \varphi^{*i } \varphi^i -M \bar{\psi}^i \psi^i -D \varphi^{*i} \varphi^i +\frac{N}{g^2} D +(\varphi^i \bar{\psi}^i\lambda +\varphi^{i*} \bar{\lambda} \psi^i)\label{lagrangian-CPN}
\eeq
Since the gauge superfield $V$ or its component fields do not have kinetic terms, they are auxiliary fields and do not propagate in the tree approximation.
If we eliminate all auxiliary fields using their equations of motion, we obtain the constraints
\beq
&&\varphi^i \varphi^{* i}=\frac{N}{g^2}, \quad 
M=-\frac{g^2}{2N} \bar{\psi}^{i}\psi^i, \nonumber\\
&&v^\mu=\frac{g^2}{2N}\left[i (\varphi^{*i} \partial_\mu \varphi^i -\varphi^i \partial_\mu \varphi^{i*})+ \bar{\psi}^{i}\gamma^\mu \psi^i\right], \label{eom-CPN}\\
&&\varphi^i \bar{\psi}^i =\varphi^{*i} \psi^i=0. \nonumber
\eeq
The first equation means fields $\varphi^i$ are constrained on the $(2N-1)$ dimensional sphere $S^{2N-1}$.
Furthermore, the gauge transformation of gauge field $v_\mu$ eliminates a common phase of $\varphi^i$.
Thus, the target manifold reduces to the complex projective space ${\bf C}P^{N-1}$. The last equation of (\ref{eom-CPN}) implies that fermion resides on the tangent space of ${\bf C}P^{N-1}$.

\subsection{Phase structure of ${\bf C}P^{N-1}$ model}
To investigate the phase structure, let us calculate the effective potential in the leading order of the $1/N$ expansion.
The partition function of this model can be written as
\beq
Z=\int D\Phi^i D \Phi^{\dag \bar{i}} DV e^{i \int d^3 x {\cal L}},
\eeq 
where the Lagrangian ${\cal L}$ is given in (\ref{CPN-kahler2}) and the measure $DV$ includes the gauge fixing term of the remaining gauge freedom in the Wess-Zumino gauge. In order to calculate the effective potential, the vacuum energy when $\langle \varphi(x)\rangle=\tilde{\varphi}$ is kept fixed,  
we divide the dynamical field into the vacuum expectation value and the fluctuation, $\varphi^i=\tilde{\varphi}^i +\varphi^{'i}$.
The fluctuation field has to follow a constraint \cite{HKNT,KH}
\beq
\int d^3 x \varphi^{'i}(x)=0,
\eeq
which forbids the appearance of tadpole or one particle reducible graphs in the effective potential.
Integrating over the fluctuation field $\varphi^{'i}$, we obtain the effective action:
\beq
Z&=&\int DV e^{i S_{eff}},\label{partition_fn}\\
S_{eff}&=&-\frac{N}{i}{\rm Tr} \ln (\nabla_{B} +\bar{\lambda} \nabla^{-1}_F \lambda) +\frac{N}{i}{\rm Tr} \ln \nabla_{F}\label{eff-action} \\
&&+\int d^3x\left(\frac{N}{g^2} D - \tilde{\varphi}^{*i }(M^2+D) \tilde{\varphi}^i+ ( \mbox{$\lambda$-dependent terms})\right). \nonumber
\eeq
Here we use the following notation
\beq
\nabla_{B}=D^\mu D_\mu +(M^2 +D),\hspace{0.5cm} \nabla_{F}=i \gamma^{\mu}D_{\mu}-M,\hspace{0.5cm} 
D_\mu = \partial_\mu +i v_\mu. \nonumber
\eeq
Assuming the Lorentz invariance of the vacuum, we take the vacuum expectation value of each auxiliary field as follow:
\beq
\langle M(x) \rangle =M_0 , \hspace{0.5cm} \langle D(x) \rangle=D_0, \hspace{0.5cm} \langle {\mbox{others}} \rangle=0. \nonumber
\eeq
We evaluate the momentum integration by introducing the ultraviolet cutoff $\Lambda$, then we obtain the effective potential in the leading order of the $1/N$ expansion ($k\llap /=\gamma^{\mu}k_{\mu}$).
\beq
\frac{V_{\rm eff}}{N}&=&\int^{\Lambda} \frac{d^3 k}{(2\pi)^3} \ln (-k^2 +M_0^2 +D_0) -\int^{\Lambda} \frac{d^3 k}{(2 \pi)^3}{\rm tr} \ln (k\llap / -M_0)\nonumber\\
&&+\frac{1}{N}\tilde{\varphi}^{i*}(M_0^2 +D_0) \tilde{\varphi}^i -\frac{1}{g^2}D_0 \nonumber\\
&=&-\frac{1}{6 \pi} |M_0^2+D_0|^{\frac{3}{2}} +\frac{1}{6\pi}|M_0|^3 +\frac{1}{N}(M_0^2 +D_0)|\tilde{\varphi}^i|^2 +(\frac{\Lambda}{2 \pi^2}-\frac{1}{g^2})D_0.\nonumber\\
&=&-\frac{1}{6 \pi} |M_0^2+D_0|^{\frac{3}{2}} +\frac{1}{6\pi}|M_0|^3 +\frac{1}{N}(M_0^2 +D_0)|\tilde{\varphi}^i|^2 +\frac{m}{4 \pi} D_0, \label{CPN-potential} 
\eeq
where we defined a renormalized coupling constant $g_R$ to absorb the linear divergence that appears in the coefficient of $D_0$
\beq
\frac{\mu}{g_R^2}&=&\frac{1}{g^2}-\frac{1}{2\pi^2}\Lambda +\frac{\mu}{4 \pi}.\label{renormalized_coupling}
\eeq
with an arbitary finite mass scale $\mu$ called the renormalization point. We have also defined a renormalization group invariant mass $m$ given by 
\beq
m &\equiv& \mu (1-\frac{4 \pi}{g_R^2}). \label{mass}
\eeq

To evaluate the effective potential in the presence of $\tilde{\varphi}, M_0, D_0$, we have to perform the path integration over the nonzero mode of auxiliary fields in (\ref{partition_fn}). When $N$ is very large, these path integration can be done by the saddle point method since the $S_{eff}$ is of order $N$. In the leading order of $1/N$ expansion, the effective potential is given by the value of $S_{eff}$ at the saddle point. Assuming the saddle point is located at the translationally invariant configuration, we can neglect the non-zero mode of the auxiliary fields in the large $N$ limit. The saddle point condition is, therefore, equivalent to the the stationary condition of the effective potential (\ref{CPN-potential}): 
\beq
\frac{1}{N} \frac{\partial V_{\rm eff}}{\partial M_0} &=&-2M_0 \Big( \frac{\epsilon}{4 \pi} |M_0^2 +D_0|^{\frac{1}{2}} -\frac{1}{4\pi}|M_0| -\frac{1}{N}|\tilde{\varphi}^i|^2 \Big) =0,\label{m0cond-CPN}\\
\frac{1}{N} \frac{\partial V_{\rm eff}}{\partial D_0}&=& -\frac{\epsilon}{4 \pi} |M_0^2 +D_0|^{\frac{1}{2}} +\frac{1}{N}|\tilde{\varphi}^i|^2 +\frac{m}{4 \pi}=0\label{d0cond-CPN}
\eeq
where $\epsilon={\rm sgn}(M_0^2 +D_0)$. These two conditions fixes the value of $M_0$ at the saddle point
\begin{equation}
|M_0|=0 \quad \mbox{or}\quad m. \label{m0_saddle}
\end{equation}
We will discuss these two cases separately.
\begin{description}
\item[$|M_0|=m$ case:]
This case is possible only when $m\ge 0$. By solving (\ref{d0cond-CPN}) for $|D_0+m^2|$ and substituting it back to the effective potential (\ref{CPN-potential}), we obtain the vacuum energy when $\tilde{\varphi}$ is kept fixed
\begin{equation}
V_{\rm eff}(\tilde{\varphi})=\frac{N}{12\pi}\left(\left|\frac{4\pi}{N}|\tilde{\varphi}^i|^2+m\right|^3-m^3\right)\label{v1_eff}
\end{equation}
Because we are assuming $m\ge 0$, the minimum of this vacuum energy is located at $\tilde{\varphi}^i=0$, (\ref{d0cond-CPN}) implies $D_0=0$. Since $\tilde{\varphi}$ and $D_0$ are the order parameter of $SU(N)$ and supersymmetry respectively, neither $SU(N)$ nor supersymmetry is broken in this case.

\item[$M_0=0$ case:]
By solving (\ref{d0cond-CPN}) for $|D_0|$ and substituting it back to the effective potential (\ref{CPN-potential}), we obtain the vacuum energy
\begin{equation}
V_{\rm eff}(\tilde{\varphi})=\frac{N}{12\pi}\left|\frac{4\pi}{N}|\tilde{\varphi}^i|^2+m\right|^3\label{v2_eff}
\end{equation}
When $m\ge 0$, the minimum located at $\tilde{\varphi}=0$ has a higher vacuum energy than the previous vacuum, and does not correspond to the ground state.
On the other hand, when $m<0$, the minimum is at 
\begin{equation}
|\tilde{\varphi}^i|^2=\frac{N}{4\pi}|m|,\label{vev_broken}
\end{equation}
then (\ref{d0cond-CPN}) implies $D_0=0$, namely supersymmetry is unbroken while $SU(N)$ symmetry is spontaneously broken in this case.
\end{description}

Thus we found two vacuum, both of them are supersymmetric. In the following, we will discuss these two phases in detail. We first discuss the mass spectrum of dynamical fields by substututing the vacuum expectation values in the lagrangian.Then we evaluate the two point functions of the auxiliary fields to discuss their masses.

\begin{enumerate}
\item $SU(N)$ symmetric phases  ($|\tilde{\varphi}^i|^2=0,  M_0 =m$)\\
By replacing $M$ by $m+M'$ in the lagrangian (\ref{lagrangian-CPN}):
\beq
{\cal L}&=&-\varphi^{*i}(\partial_\mu \partial^\mu+m^2)\varphi^i+\bar{\psi}^i(i{\partial}\llap / -m)\psi^i+F^i F^{*i} \nonumber\\
&&-[i (\varphi^{* i } \partial_\mu \varphi^i -\varphi^i \partial_\mu \varphi^{*i}) +\bar{\psi}^i \gamma_\mu \psi^i ]v^\mu +v^\mu v_\mu \varphi^{*i } \varphi^i 
\label{shifted_lagrangian-CPN}\\
&&- (2mM'+M^{'2}) \varphi^{*i } \varphi^i -M' \bar{\psi}^i \psi^i -D \varphi^{*i} \varphi^i +\frac{N}{g^2} D +(\varphi^i \bar{\psi}^i \lambda +\varphi^{*i} \bar{\lambda} \psi^i),\nonumber
\eeq
we find that $N$ scalar and spinor fields have the same mass $m$ by the vacuum expectation value of $M_0$, and global SU(N) and local $U(1)$ gauge symmetry are unbroken in this phase.
Integrating out the dynamical field, we obtain the effective action (\ref{eff-action}) describing the dynamics of auxiliary fields in this phase.
\beq
S_{eff}&=&-\frac{N}{i}{\rm Tr}\ln \{ \partial_\mu \partial^\mu +i (v^\mu \partial_\mu +\partial^\mu v_{\mu})-v^\mu v_\mu + \bar{\lambda} [i {\partial}\llap / -v\llap /-(m+M')]^{-1}\lambda \nonumber\\
&&+[(m+M')^2 +D]  \}+\frac{N}{i}{\rm Tr} \ln [i {\partial}\llap / -v\llap / -(m+M')] +\int dx \frac{N}{g^2}D.
    \eeq
When we expand this $S_{eff}$ in the power series of fields, bilinear terms define the propagator of the auxiliary fields, and higher order terms define the interaction terms of auxiliary fields. 
Two point function of all auxiliary fields are summarized in Appendix.\ref{feynman-CPN}, there we find the mass of the fields $v_{\mu}, M, \lambda$ is twice of that of the dynamical fields. These auxiliary field $v_{\mu}, M$ and $\lambda$ represent the bound states of fermion--anti-fermion and fermion--boson as is shown in (\ref{eom-CPN}). 

\item $SU(N)$ broken phases  ($|\tilde{\varphi}^i|^2=\frac{N}{4 \pi}|m|,  M_0 =0$)\\
Using $SU(N)$ symmetry, we choose the vacuum expectation value of $\tilde{\varphi}$ in the $N$-th direction: $\tilde{\varphi}^N=\sqrt{\frac{N}{4 \pi}|m|}$. Substituting $\varphi^N=\sqrt{\frac{N}{4 \pi}|m|}+\varphi'{}^{N}$ to (\ref{lagrangian-CPN}), we obtain
\begin{eqnarray}
{\cal L}&=&{\cal L}_{N-1}+{\cal L}_N\nonumber\\
{\cal L}_{N-1}&=&\sum_{i=1}^{N-1}\left[-\varphi^{i*}(D_\mu D^\mu+M^2+D)\varphi^i+\bar{\psi}^i(i\gamma^{\mu}D_{\mu}-M)\psi^i +F^i F^{i*}+(\varphi^i \bar{\psi}^i \lambda +\varphi^{i*}\bar{\lambda} \psi^i)\right]\nonumber\\
{\cal L}_N&=&-\varphi'{}^{N*}(D_\mu D^\mu+M^2+D)\varphi'{}^N+\bar{\psi}^N(i\gamma^{\mu}D_{\mu}-M)\psi^N +F^N F^{N*}+\frac{N}{g^2} D \nonumber\\
&&+(\varphi'{}^N \bar{\psi}^N \lambda +\varphi'{}^{N*}\bar{\lambda} \psi^N)
\nonumber\\
&&+\sqrt{\frac{N}{4\pi}|m|} \left[-D(\varphi'{}^N+ \varphi'{}^{N*})-iv^\mu \partial_\mu (\varphi'{}^N - \varphi'{}^{N*})+(\bar{\psi}^N \lambda +\bar{\lambda}\psi^N )\right]\label{mixing_term}\\
&&-\sqrt{\frac{N}{4\pi}|m|}(M^2-v^\mu v_\mu)(\varphi'{}^N + \varphi'{}^{N*})
-\frac{N}{4 \pi}|m|(M^2-v^\mu v_\mu) \nonumber
\end{eqnarray}
In this phase, global $SU(N)$ symmetry is broken down to $SU(N-1)$ and there are $N-1$ massless Nambu-Goldstone bosons $\varphi^i\ (i=1,\cdots,N-1)$, and their superpartners $\psi^i\ (i=1,\cdots,N-1)$.
Furthermore, gauged $U(1)$ symmetry is also broken because the dynamical fields $\varphi^N$, carrying $U(1)$ charge, have a nonvanishing vacuum expectation value.
The third line of ${\cal L}_N$ represents the mixing between $\varphi'{}^N, \psi^N$ and auxiliary fields.
Integrating over $\varphi^i, \psi^i$ with $i=1,\cdots,N$ in (\ref{partition_fn}), we obtain the effective action for auxiliary fields:
\beq
S_{eff}&=&-\frac{N-1}{i}{\rm Tr}\ln \{ \partial^\mu \partial_\mu +i (v^\mu \partial_\mu +\partial^\mu v_\mu ) -v^\mu v_\mu +\bar{\lambda} [i \partial\llap / -v\llap / -M ]^{-1} \lambda +[M^2 +D] \}\nonumber\\
&&+\frac{N-1}{i}{\rm Tr}\ln [i \partial\llap / -v\llap / -M]
+\int dx \frac{N}{g^2}D\label{CP-broken-action}.
\eeq
In order to obtain propagators of auxiliary fields we have to take into account the mixing between auxiliary fields and $\varphi'{}^N, \psi^N$. To diagonalize the propagators, we add the quadratic terms in (\ref{mixing_term}) to the bilinear terms of $S_{eff}$ to define $S^{(2)}_{eff}$
\beq
S^{(2)}_{eff}&=&\int \frac{d^3 p }{(2\pi)^3} 
\Bigg( \frac{1}{2}M(-p) \Big(-\frac{N}{8}\sqrt{-p^2} -\frac{N}{2\pi}|m| \Big)M(p) 
-\frac{1}{2} D(-p) \frac{N}{8 \sqrt{-p^2}} D(p) \nonumber\\
&&-\frac{1}{2} v^\mu (-p) \Big( \frac{N}{8} \frac{p^2 \eta_{\mu \nu} -p_\mu p_\nu}{\sqrt{-p^2}} -\frac{N}{2\pi} |m| \eta_{\mu \nu} \Big) v^\nu (p)
+\bar{\lambda} (-p) \frac{N p\llap /}{16\sqrt{-p^2}} \lambda (p) \nonumber\\
&&+\frac{1}{2} \varphi^N_R(-p) p^2 \varphi^N_R(p) + \frac{1}{2} \varphi^N_I(p) p^2 \varphi^N_I (p)+\bar{\psi}^N (-p) p\llap / \psi^N (p) \label{s_eff-CPNbroken}\\
&&-\frac{1}{2}\sqrt{\frac{N}{2\pi} |m|} (\varphi^N_R D(p) +D(-p) \varphi^N_R )
+\frac{1}{2} \sqrt{\frac{N}{2 \pi} |m|} (\bar{\psi}^N (-p) \lambda(p) +\bar{\lambda}(-p)\psi^N (p) )\nonumber\\
&&-\frac{1}{2}i p^\mu \sqrt{\frac{N}{2\pi}|m|} (v_\mu (-p) \varphi^N_I (p) - \varphi^N_I (-p) v_\mu (p)) \Bigg),\nonumber
\eeq
where $\varphi'{}^N= \frac{1}{\sqrt{2}}(\varphi^N_R +i \varphi^N_I)$.
We immediately find that the propagator of the auxiliary field $M$ has no pole.
To solve the mixing between the gauge field $v^\mu$ and $\varphi^N_I$, we define a new vector field $U^\mu (p) = v^\mu (p)-i p^\mu (\sqrt{\frac{N}{2\pi}|m|})^{-1} \varphi^N_I$ and obtain its propagator:
\beq
\Pi^{-1}_{\mu \nu}=\frac{1}{N} \frac{1}{-\frac{1}{8} \sqrt{-p^2} -\frac{1}{2\pi}|m| } \Big( \eta_{\mu \nu} -\frac{2 \pi}{|m|} \frac{p_\mu p_\nu}{8 \sqrt{-p^2}} \Big).
\eeq
The singularity of this propagator is located at
\beq
-\frac{1}{8}\sqrt{-p^2} -\frac{1}{2 \pi}|m|=0,
\eeq
This equation has no solution in the physical sheet: $-\pi\le {\rm arg}(-p^2)\le \pi$.
Similarly $D, \lambda$ mix with $\varphi^N_R, \psi^N$, respectively, 
Then the eigen values of the field propagators are calculated as follow:
	\begin{align}
\frac{1}{2}
\begin{pmatrix}
 D(-p) & \varphi^N_{R}(-p) 
\end{pmatrix} \!\!
\begin{pmatrix}
 \frac{N}{8 \sqrt{-p^2}} & - \sqrt{\frac{N}{2 \pi}|m|} \\
- \sqrt{\frac{N}{2 \pi}|m|} & p^2
\end{pmatrix} \!\!
\begin{pmatrix}
D(p) \\
\varphi^N_R(p) 
\end{pmatrix}
&= \frac{1}{2}
\begin{pmatrix}
D(-p) & \varphi^N_{R}(-p) 
\end{pmatrix}
G^{-1}
\begin{pmatrix} 
D(p) \\
\varphi^N_R(p) 
\end{pmatrix}, \nonumber \\
\begin{pmatrix}
  \bar{\lambda}(-p) & \bar{\psi}^N(-p)
\end{pmatrix} \!\!
\begin{pmatrix}
 \frac{N p\llap /}{16\sqrt{-p^2}} & \sqrt{\frac{N}{4 \pi}|m|} \\
 \sqrt{\frac{N}{4 \pi}|m|} &  p\llap /
\end{pmatrix} \!\!
\begin{pmatrix}
\lambda(p) \\
\psi^N(p)
\end{pmatrix}
&=
\begin{pmatrix}
  \bar{\lambda}(-p) & \bar{\psi}^N(-p)
\end{pmatrix} 
S^{-1}
\begin{pmatrix} 
\lambda(p) \\
\psi^N(p)
\end{pmatrix}.
\end{align}
The poles of the new propagators ($G$ and $S$) are obtained by solving
\beq
\det G^{-1} &=&-\frac{N}{8}\sqrt{-p^2} -\frac{N}{2\pi} |m|=0, \\
\det S^{-1} &=&-\frac{N}{16}\sqrt{-p^2} -\frac{N}{4\pi} |m|=0.
\eeq
However these equations have no solution in the physical Riemann sheet.
Thus we find all auxiliary fields are not physical particles in the broken phase.
\end{enumerate}

\subsection{The $\beta$ function of $CP^{N-1}$}
To derive $\beta$ function, we calculate the next-to-leading correction of the gap equation in the symmetric phase.
The $\beta$ function is determined by the UV behavior and we should be able to define a universal $\beta$ function in both phases.
Feynman rules in symmetric phase are summarized in Appendix \ref{feynman-CPN}.
In our notation, propagators of auxiliary field have the factor $1/N$.

The gap equation is given by equation of motion of the auxiliary field $D$.
\beq
\frac{N}{g^2}=\int \frac{d^3 p}{(2 \pi)^3} \langle \varphi^i (p) \varphi^{*i} (0) \rangle. \label{gap-cpn}
\eeq
In the leading order of $1/N$, the gap equation is written as follow:
\beq
\frac{N}{g^2}=\int \frac{d^3 p}{(2 \pi)^3} \frac{i N}{p^2-m^2}.
\eeq
We renormalize the coupling constant to absorb the divergence using UV cutoff and obtain the leading order $\beta$ function of $1/N$ expansion.
\beq
\beta(g_R)=-\frac{1}{8 \pi} g_R^3 +\frac{1}{2}g_R \label{beta-cpn}
\eeq

\begin{figure}[h]
\begin{center}

\unitlength=1mm
\begin{picture}(50,40)

\includegraphics[width=5cm]{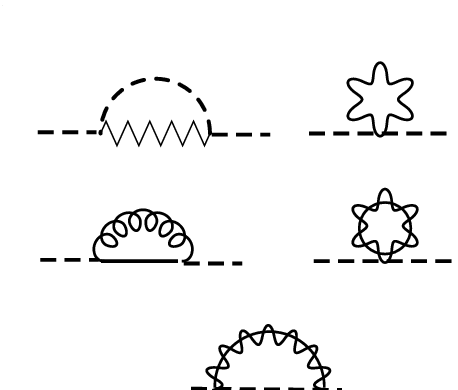}
\put(-51,24){\bf $\varphi^i$}
\put(-24,24){\bf $\varphi^i$}
\put(-35,24){\bf $\alpha$}

\put(-18,24){\bf $\varphi^i$}
\put(2,24){\bf $\varphi^i$}
\put(-4,33){\bf $\sigma$}

\put(-51,13){\bf $\varphi^i$}
\put(-24,13){\bf $\varphi^i$}
\put(-37,10){\bf $\psi^i$}
\put(-29,18){\bf $\xi$}

\put(-20,13){\bf $\varphi^i$}
\put(0,13){\bf $\varphi^i$}
\put(-4,18){\bf $v^\mu$}

\put(-30,-3){\bf $\varphi^i$}
\put(-15,-3){\bf $\varphi^i$}
\put(-12,2){\bf $v^\mu$}

\end{picture}
\caption{The Feynman diagrams contributing to the next-to-leading order correction of $\varphi$ propagator}\label{cpn-next}

\end{center}

\end{figure}

We calculate the next-to-leading Feynman diagrams, shown Fig.\ref{cpn-next}, and find the next-to-leading correction of $\beta$ function vanishes because of ${\cal N}=2$ supersymmetry.

\begin{figure}[h]
\begin{center}

\unitlength=1mm
\begin{picture}(50,40)

\includegraphics[width=5cm]{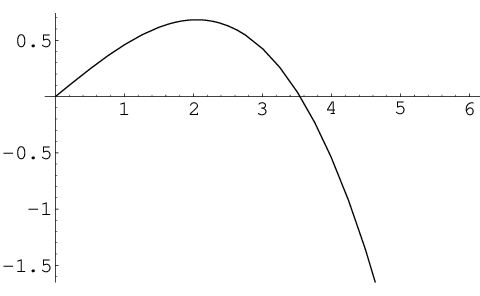}
\put(-51,30){\bf $\beta (g)$}
\put(-20,17) {\bf$g_c$}
\put(0,17){\bf $g$}
\put(-15,24){\bf The symmetric and}
\put(5,20) {\bf massive phase}
\put(-42,0){\bf The broken and} 
\put(-42,-4){\bf massless phase}

\end{picture}
\caption{The $\beta$ function of the coupling constant $g$ for the $CP^{N-1}$ model.}\label{3dim-beta}

\end{center}

\end{figure}
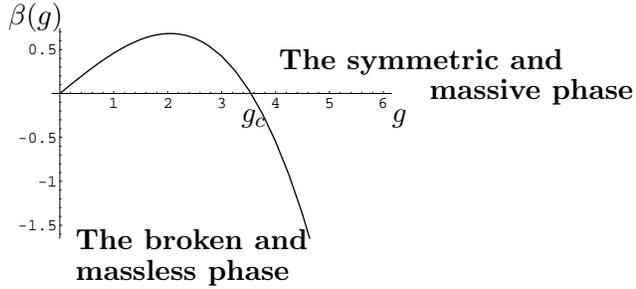

Finally, let's compare the result with the WRG result of $CP^{N-1}$ case \cite{HI-3dim}:
\beq
\beta (\lambda)=-\frac{N}{4 \pi^2} \lambda^3+\frac{1}{2}\lambda.
\eeq
We define the 't Hooft coupling $g^2 \equiv N\lambda^2$, and find both $\beta$ functions are same.
This $\beta$ function is shown in Fig.\ref{3dim-beta}.
From the eq.(\ref{mass}), the region where the coupling constant is smaller than the critical point $g_c$ corresponds to broken phase.

\section{The $Q^{N-2}$ model}\label{sec-QN}
\subsection{The theory and symmetry}
The $\beta$ function for the $CP^{N-1}$ model has not next-to-leading order correction because of ${\cal N}=2$ supersymmetry.
The question is that all ${\cal N}=2$ supersymmetric nonlinear sigma models have no next-to-leading corrections in terms of $1/N$ expansion.
To investigate it, we consider the other ${\cal N}=2$ supersymmtric nonlinear sigma model; the $Q^{N-2}$ model.
The Wilsonian renormalization group (WRG) approach for the $Q^{N-2}$ model gives the following $\beta$ function:
\beq
\beta (\lambda)&=&-\frac{N-2}{4 \pi^2} \lambda^3+\frac{1}{2}\lambda, \nonumber
\eeq
This equation can be rewritten using 't Hooft coupling $g^2=N\lambda^2$,
\beq
\beta(g)&=&-\frac{g^3}{4 \pi^2}(1-\frac{2}{N})+\frac{1}{2}g.
\eeq
This WRG result shows this model has the next-to-leading correction of $\beta$ function in $1/N$ expansion.

Let's investigate $Q^{N-2}$ model using $1/N$ expansion.
This model is obtained from $CP^{N-1}$ model with the $O(N)$ condition $\Phi^i \Phi^i =0$.
Then the Lagrangian has the two kinds of auxiliary fields as follow:
\beq
{\cal L}=\int d^4 \theta (\Phi^i \Phi^{\dag \bar{i}} e^V -c V)+\frac{1}{2} \Big( \int d^2 \theta \Phi_0 \Phi^i \Phi^i +\int d^2 \bar{\theta} \Phi^\dag_0 \Phi^{\dag i } \Phi^{\dag i}   \Big),
\eeq
where $V(\theta,\bar{\theta},x)$ is $U(1)$ gauge superfield and $\Phi_0 ,\Phi^{\dag}_0$ are chiral- and antichiral- superfields respectively.
Components of the new auxiliary field $\Phi_0$ is defined by
\beq
\Phi_0(y)&=&A_0 (y)+\sqrt{2} \theta \psi^i_0 (y) +\theta \theta F_0 (y).\nonumber
\eeq
This auxiliary superfield is an $O(N)$ singlet and has gauged nonzero $U(1)$ charge; $[\Phi_0]=-2$ and $[\Phi_0^\dag]=2$.
The dynamical fields $\Phi^i$ is a vector representation of global $O(N)$ symmetry and has the gauged $U(1)$ charge (\ref{U(1)sym}).
Then the $Q^{N-2}$ model has global $SO(N)$ symmetry coming from the isometry of the target manifold and the gauged $U(1)$ symmetry.

Integrated out the Grassmann coordinates $\theta$ and $\bar{\theta}$, we obtain the $Q^{N-2}$ model Lagrangian in component fields:
\beq
{\cal L}&=&\partial_\mu \varphi^{*i } \partial^\mu \varphi^i +i \bar{\psi} \partial_\mu \gamma^\mu \psi^i +F^i F^{*i} -[i (\varphi^{*i } \partial_\mu \varphi^i -\varphi^i \partial_\mu \varphi^{*i}) -\bar{\psi}^i \gamma_\mu \psi^i ]v^\mu +v^\mu v_\mu \varphi^{*i } \varphi^i \nonumber\\
&& - M^2 \varphi^{*i } \varphi^i -M \bar{\psi}^i \psi^i -D \varphi^{*i} \varphi^i +\frac{N}{g^2} D +(\varphi^i \bar{\psi}^i \lambda ^c +\varphi^{*i} \lambda^c \psi^i)\nonumber\\
&&+\frac{1}{2}(F_0 \varphi^i \varphi^i +F_0^{*} \varphi^{* i} \varphi^{* i})-(\bar{\psi}_0^c \psi^i \varphi^i +\bar{\psi}^i \psi_0^c \varphi^{* i}) \nonumber\\
&&+A_0 (\varphi^i F^i -\frac{1}{2} \bar{\psi}^{c i} \psi^i)
+A_0^*(\varphi^{*i} F^{*i} -\frac{1}{2} \bar{\psi}^{ i} \psi^{c i}).
\eeq
If we eliminate auxiliary fields using their equation of motion, we obtain the constraint eqs.(\ref{eom-CPN}) and additional following equations.
\beq
\varphi^i \varphi^i=\varphi^{*i} \varphi^{*i}=0, \hspace{0.5cm} 
\varphi^i \psi^i=\varphi^{*i} \bar{\psi}^{i}=0, \hspace{0.5cm}
A_0=-\frac{g^2}{2N} \bar{\psi}^{c i}\psi^{i}, \hspace{0.5cm}
A_0^{*}=-\frac{g^2}{2N} \bar{\psi}^{ i}\psi^{ci}.\label{eom-QN}
\eeq
The equations (\ref{eom-QN}) is similar to ${\cal N}=1$ supersymmetric $O(N)$ model with zero radius.

\subsection{Phase structure}

Now we obtain the stationary conditions of the effective potential.

Similarly to $CP^{N-1}$ case, we integrate out the fluctuation field and put the vacuum expectation values as follows:
\beq
\langle M(x) \rangle &=&M_0, \hspace{0.5cm} \langle D(x) \rangle=D_0,  \nonumber\\
\langle F_0(x) \rangle &=&F_0 , \hspace{0.5cm} \langle A_0(x) \rangle=A_0, \hspace{0.5cm} \langle \mbox{the others} \rangle=0. \nonumber
\eeq
Then we obtain the effective potential.
\beq
\frac{V}{N}&=&\int \frac{d^3 k}{(2\pi)^3} \ln (k_\mu k^\mu +M_0^2 +D_0+|A_0|^2-|F_0|^2) -\int \frac{d^3 k}{(2 \pi)^3}tr \ln (-k^\mu \gamma_\mu -M_0-|A_0|^2)\nonumber\\
&&+\frac{1}{N}\tilde{\varphi}^{i*}(M_0^2 +D_0+A_0 A_0^*) \tilde{\varphi}^i -\frac{1}{g^2}D_0 -\frac{1}{2N}(F_0 \varphi^i \varphi^i +F_0^* \varphi^{*i}\varphi^{*i}) \nonumber\\
&=&-\frac{1}{6 \pi} \frac{1}{2} \Big[|M_0^2+D_0+|A_0|^2 +|F_0||^{\frac{3}{2}}+|M_0^2+D_0+|A_0|^2 -|F_0||^{\frac{3}{2}} \Big]\nonumber\\
&& +\frac{1}{6\pi}\frac{1}{2}\Big[ |M_0+|A_0||^3 +|M_0-|A_0||^3 \Big] -(\frac{1}{g^2} -\frac{1}{2\pi^2}\Lambda)D_0 \nonumber\\
&&+\frac{1}{N} \Big(M_0^2 \varphi^{*i} \varphi^i +A_0^* A_0 \varphi^{*i}\varphi^i +D_0 \varphi^{*i}\varphi^i -\frac{1}{2} (F_0 \varphi^i \varphi^i +F_0^* \varphi^{*i}\varphi^{*i}) \Big),\nonumber\\
&=&-\frac{1}{6 \pi} \frac{1}{2} \Big[|M_0^2+D_0+|A_0|^2 +|F_0||^{\frac{3}{2}}+|M_0^2+D_0+|A_0|^2 -|F_0||^{\frac{3}{2}} \Big]\nonumber\\
&& +\frac{1}{6\pi}\frac{1}{2}\Big[ |M_0+|A_0||^3 +|M_0-|A_0||^3 \Big] +\frac{m}{4 \pi}D_0 \nonumber\\
&&+\frac{1}{N} \Big(M_0^2 \varphi^{*i} \varphi^i +A_0^* A_0 \varphi^{*i}\varphi^i +D_0 \varphi^{*i}\varphi^i -\frac{1}{2} (F_0 \varphi^i \varphi^i +F_0^* \varphi^{*i}\varphi^{*i}) \Big).\nonumber
\eeq
We use the same renormalization of coupling constant and the same invariant mass as $CP^{N-1}$ case.

If the all contents of absolute value are positive, the stationary points satisfy the following equations.
\beq
\frac{1}{N} \frac{\partial V}{\partial M_1} &=&-\frac{1}{4 \pi} M_1 \Big[ |M_1^2+M_2^2 +D_1|^{\frac{1}{2}} +|M_1^2 +M_2^2 +D_2|^{\frac{1}{2}} +2 \sqrt{2} |M_1|\Big] \nonumber\\
&&+\frac{1}{N} M_1 (\varphi_1^2 +\varphi_2^2)=0\nonumber\\
\frac{1}{N} \frac{\partial V}{\partial M_2} &=&-\frac{1}{4 \pi} M_1 \Big[ |M_1^2+M_2^2 +D_1|^{\frac{1}{2}} +|M_1^2 +M_2^2 +D_2|^{\frac{1}{2}} +2 \sqrt{2} |M_2|\Big] \nonumber\\
&&+\frac{1}{N} M_2 (\varphi_1^2 +\varphi_2^2)=0\nonumber\\
\frac{1}{N} \frac{\partial V}{\partial D} &=&-\frac{1}{8 \pi} \Big[ |M_1^2+M_2^2 +D_1|^{\frac{1}{2}} +|M_1^2 +M_2^2 +D_2|^{\frac{1}{2}} 
\frac{m}{8 \pi}+\frac{1}{2N} (\varphi_1^2 +\varphi_2^2)=0\nonumber\\
\frac{1}{N} \frac{\partial V}{\partial F_0} &=&\frac{1}{16 \pi} \Big[e^{-2i \theta}|M_1^2 +M_2^2+D_1|^{\frac{1}{2}}-e^{-2i \theta}|M_1^2+M_2^2+D_2|^{\frac{1}{2}} \Big] -\frac{1}{2N}\varphi^i \varphi^i=0\nonumber\\
\frac{1}{N} \frac{\partial V}{\partial F_0^*} &=&\frac{1}{16 \pi} \Big[e^{2i \theta}|M_1^2 +M_2^2+D_1|^{\frac{1}{2}}-e^{2i \theta}|M_1^2+M_2^2+D_2|^{\frac{1}{2}} \Big] -\frac{1}{2N}\varphi^{*i} \varphi^{*i}=0\nonumber
\eeq
where we defined new parameters as follow:
\beq
F_0&=&|F_0|e^{2i \theta}, \hspace{0.5cm} \frac{1}{\sqrt{2}}(M + |A_0|)=M_1, 
\hspace{0.5cm} \frac{1}{\sqrt{2}}(M - |A_0|)=M_2, \nonumber\\
D-|F_0|&=&D_1, \hspace{0.5cm} D+|F_0|=D_2, \hspace{0.5cm} \varphi^i e^{i \theta}=\frac{1}{\sqrt{2}}(\varphi_1+i \varphi_2). \nonumber
\eeq

We find three kinds of phases.
\begin{enumerate}
	\item Chern-Simons phases
	
	The case of $M_0=m$, $|A_0|=0$, $D=F_0=\varphi_1=\varphi_2=0$.
	
	In this phase, both global $SO(N)$ and gauged $U(1)$ symmetries are preserved.
	$N$ dynamical scalar and spinor fields obtain masses $m$ due to the vacuum expectation value of $M_0$.
	Similarly to $CP^{N-1}$ case, the auxiliary fields have twice masses of the dynamical fields.
	The auxiliary field with the vacuum expectation value $M$ corresponds to the bound state of the fermion and anti-fermion.
	
	Expanding the effective action, the gauge field parts of the effective action have the Chern-Simons interaction term in the next-to-leading order of $1/N$.
	\beq
	S_{eff} \sim \int \frac{d^3 p}{(2 \pi)^3} - v^\mu (-p) \frac{N}{4\pi} (p^2 \eta_{\mu \nu} -p_\mu p_\nu -2mi \epsilon_{\mu \nu \rho}p^{\rho}) I^{-1}(p) v^\mu,
	\eeq
	where, 
	\beq
	I(p)=\frac{\sqrt{-p^2}}{\arctan({\sqrt{\frac{-p^2}{4m}}})}.
	\eeq
	The gauge field $v^\mu$ obtain the mass $2m$ becuase of this interaction term in Chern- Simons phase.
	
	\item Higgs phases
	The case of $M_0=0$, $|A_0|=m$, $D=F_0=\varphi_1=\varphi_2=0$.
	
	In this phase, global $SO(N)$ symmetry is protected and $N$ dynamical fields have masses due to the vacuum expectation value of $A_0$.
	However gauged $U(1)$ symmetry is broken because the superfields $\Phi_0$, which has nonzero $U(1)$ charge, has nonvanishing vacuum expectation value.

	Expanding the effective action, we calculate all propagators of auxiliary fields in next-to-leading order of $1/N$:
	\beq
	G^{A_R}&=&\frac{4 \pi}{N} \frac{i}{p^2 -4m^2}I(p),\nonumber\\
	G^{M}&=&\frac{4 \pi}{N} \frac{i}{p^2 -4m^2}I(p),\nonumber\\
	G^{D'}&=&\frac{4 \pi i}{N} I(p),\nonumber\\
	G^{F_0}&=&\frac{8 \pi}{N} I(p),\nonumber\\
	G^{\lambda'}&=&\frac{4 \pi}{N} \frac{2i}{p^\mu \gamma_\mu +2m}I(p),\nonumber\\
	G^{\psi'_0}&=&\frac{4 \pi}{N} \frac{2i}{p^\mu \gamma_\mu -2m}I(p),\nonumber\\
	G^{U}_{\mu \nu}&=&-\frac{4 \pi}{N} \frac{i}{p^2 -4m^2}(\eta_{\mu \nu}-\frac{p_\mu p_\nu}{4m^2})I(p),\nonumber
	\eeq
	where we combinate the fields $\lambda$, $\psi$, $D$, $A_0$ and $v_\mu$ to diagonalize the two point functions as follow:
	\beq
	\lambda' &=&\frac{1}{\sqrt{2}}(\lambda -\psi_0^c ),\nonumber\\
	\psi'_0 &=&\frac{i}{\sqrt{2}}(\lambda^c +\psi_0 ),\nonumber\\
	A_0&=&m+ A_R+i A_I,\nonumber\\
	D'&=&D+2mA_R,\nonumber\\
	U_\mu (p)&=&v_\mu (p) -(2m)^{-1} i p_\mu A_I. (p)\nonumber
	\eeq
	From these propagators, we find the auxiliary fields $M$ and $A_R$ has twice masses of dynamical fields.
	From eq. of motion for $A_0$, we find the fields $A_0^*$ corresponds to the pair of fermion and fermion:
	\beq
	A_0^*=\frac{g^2}{4N} \bar{\psi}^{ci} \psi^i.
	\eeq 
	In this phase, the gauge bosons acquire masses through the Higgs mechanism and the imaginary part of $A_0$ is removed from the theory.

	\item broken phases 
	
	The case of $M_0=0$, $|A_0|=0$, $D=F_0=0$, $0=\frac{4 \pi}{N}\varphi_1^2+m$, $0=\frac{4 \pi}{N}\varphi_2^2 +m$ ($m<0$).
	
	In this phase, both global $SO(N)$ and gauged $U(1)$ symmetries are broken.
	There are $N-2$ massless Nambu-Goldstone bosons and their superpartners.
	Similarly to the broken phase of $CP^{N-1}$ model, the propagators of the fields $v^\mu$, $M$, $\lambda$, $D$, $A_0$, $\psi_0$, $F_0$, $\varphi^{N-1}$, $\varphi^{N}$, $\psi^{N-1}$ and $\psi^{N}$ do not have the singular point, shown Appendix \ref{QN-broken}.
	We find these fields are not physical particles.

\end{enumerate}

\subsection{The $\beta$ function}
Finally, we calculate the $\beta$ function in Chern-Simon phase because the $\beta$ function is determined by the UV behavior and gives same result in either phase.
We summarize the Feynman rule in Chern-Simon phase in Appendix \ref{feynman-QN}.

Similarly to $CP^{N-1}$ model, the gap equation is given by
\beq
\frac{N}{g^2}=\int \frac{d^3 p}{(2 \pi)^3} \langle \varphi^i (p) \varphi^{*i} (0) \rangle. \label{gap-qn}
\eeq
The gap equation of the leading order of $1/N$ expansion is
\beq
\frac{N}{g^2}=\int \frac{d^3 p}{(2 \pi)^3} \frac{i N}{p^2-m^2}.
\eeq
Then the $\beta$ function of the coupling constant is given by
\beq
\beta(g_R)=-\frac{1}{8 \pi} g_R^3 +\frac{1}{2}g_R.
\eeq

\begin{figure}[h]
\begin{center}

\unitlength=1mm
\begin{picture}(50,40)

\includegraphics[width=5cm]{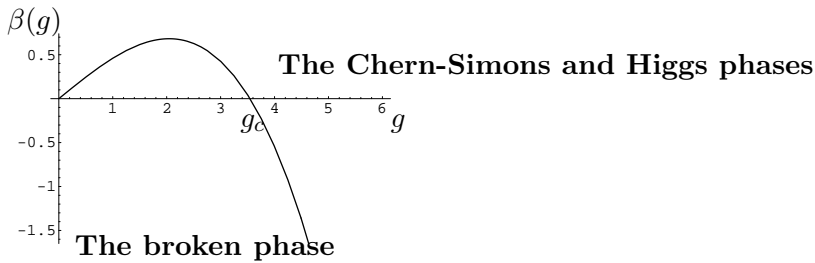}
\put(-51,30){\bf $\beta (g)$}
\put(-20,17) {\bf$g_c$}
\put(0,17){\bf $g$}
\put(-15,24){\bf The Chern-Simons and Higgs phases}
\put(-42,0){\bf The broken phase}

\end{picture}
\caption{The $\beta$ function of the coupling constant $g$ for $Q^{N-2}$ model.}\label{QN-beta-fig}

\end{center}

\end{figure}

\begin{figure}[h]
\begin{center}

\unitlength=1mm
\begin{picture}(40,40)

\includegraphics[width=5cm]{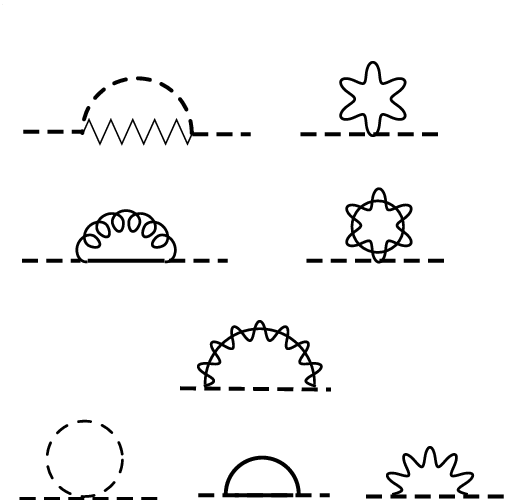}
\put(-51,35){\bf $\varphi^i$}
\put(-26,35){\bf $\varphi^i$}
\put(-38,32){\bf $D$}

\put(-10,38){\bf $M$}

\put(-40,20){\bf $\psi^i$}
\put(-31,26){\bf $\lambda$}

\put(-9,27){\bf $v^\mu$}

\put(-18,12){\bf $v^\mu$}

\put(-38,4){\bf $A_0$}

\put(-24,4){\bf $\psi_0$}
\put(-26,-4){\bf $\psi^i$}

\put(-2,4){\bf $F_0$}

\end{picture}
\caption{The Feynman diagrams contributing to the next-to-leading order correction of $\varphi$ propagetor}\label{qn-next}

\end{center}

\end{figure}

In the next-to-leading order of $1/N$, we calculate eight Feynman diagrams in Fig.\ref{qn-next}.
We introduce the UV cutoff $\Lambda$ then the gap equation is given by
\beq
\frac{1}{g^2}&=&\frac{1}{2 \pi^2} \Big( (\Lambda-\frac{\pi}{2}m) -\frac{2}{N} (\Lambda -\frac{\pi}{2} 2m) \Big) \nonumber\\
&=&\frac{\Lambda}{2 \pi}(1-\frac{2}{N}) -\frac{m}{4 \pi} (1-\frac{4}{N}).
\eeq
In three dimensional $Q^{N-2}$ model, the coupling constant $g$ has the canonical dimension $\frac{1}{2}$.
Then the $\beta$ function of such coupling constant depends on the subtraction of finite quantity \cite{tHooft}.
We define the renormalized coupling constant $g_{R}$ by
\beq
\frac{1}{g^2}-(1-\frac{2}{N})\frac{\Lambda -\mu}{2 \pi^2} =\frac{\mu}{g_R^2},
\eeq
and obtain the next-to-leading order $\beta$ function.
\beq
\beta=\frac{1}{2}g_R -\frac{1}{4 \pi^2} (1-\frac{2}{N})g_R^3
\eeq
We find this $\beta$ function coincides with the WRG result and is shown in Fig.\ref{QN-beta-fig}.

Similarly to the $CP^{N-1}$ model, the region where the coupling constant is smaller than the critical point $g_c$ corresponds to broken phase, while the region where the coupling constant is larger than $g_c$ corresponds to Chern-Simons and Higgs phase.

\section{Conclusion}
In this paper, we discuss two concrete ${\cal N}=2$ supersymmetric nonlinear sigma models, $CP^{N-1}$ and $Q^{N-2}$ models, using $1/N$ expansion.

We derived the $\beta$ function and found the non-trivial UV fixed points of these two model.
In the $CP^{N-1}$ case, the $\beta$ function accidentally has no next-to-leading order correction in the $1/N$, however in the case of $Q^{N-2}$ model, there is a next-to-leading order correction.
We found these $\beta$ functions up to next-to-leading order of $1/N$ expansion coincide with the ones obtained by using the WRG analysis.

Furthermore, we investigated the phase structures of both $CP^{N-1}$ and $Q^{N-2}$ models.

The $CP^{N-1}$ model has two phases: the $SU(N)$ symmetric and broken phases.
In the symmetric phase, all dynamical fields have mass ($m$) due to the vacuum expectation value of the auxiliary field $M$.
The auxiliary fields also have a mass ($2m$), and the field $M$ corresponds to the bound state of two dynamical spinor field $\bar{\psi}^i \psi^i$.
On the other hand, in the broken phase, a dynamical field has a vacuum expectation value, and global $SU(N)$ symmetry is broken.
There are $(N-1)$ massless Nambu-Goldstone bosons and their superpartners, and non of the auxiliary fields are physical particles.

The $Q^{N-2}$ model has three phases: the Chern-Simons, Higgs and $SO(N)$ broken phases.
In the Chern-Simons and Higgs phase, all dynamical fields have a mass ($m$) due to the vacuum expectation value of the auxiliary field $M$ and $A_0$, respectively.
The auxiliary fields also have a mass ($2m$), and the field $M$ corresponds to the bound state of the fermion and anti-fermion; $M \sim \bar{\psi}^i \psi^i$, and $A_0$ corresponds to the bound state of two fermions $A_0 \sim \bar{\psi}^{ci} \psi^i$.
In the Chern-Simons phase, the gauge fields obtain the mass due to the induced Chern-Simons interaction.
In the Higgs phase, the gauge field acquire the mass through the Higgs mechanism and the imaginary part of the field $A_0$ is absorbed by the gauge field.
In the broken phase, two components of dynamical fields  $\varphi^{N-1}$ and $\varphi^N$ have the vacuum expectation values, then global $SO(N)$ symmetry is broken down to $SO(N-2)$.
There are $(N-2)$ massless Nambu-Goldstone bosons and their superpartners, and non of the auxiliary fields are physical particles.

\section*{Acknowledgements}
We would like to thank Takeo Inami and Kunio Hitotsumatsu for useful discussions. 
This work was supported in part by the Grant-in-Aid for Scientific
Research (\#16340075 and \#13135215).
E.I is supported by Research Fellowship of the Japan Society for the Promotion of Science (JSPS) for Young Scientists (No.16-07971).

\newpage
\appendix
\section{The Feynman rules}

We summarize the Feynman rules of ${\bf C}P^{N-1}$ and $Q^{N-2}$ models in symmetric phase.
\subsection{The ${\bf C}P^{N-1}$ model}\label{feynman-CPN}

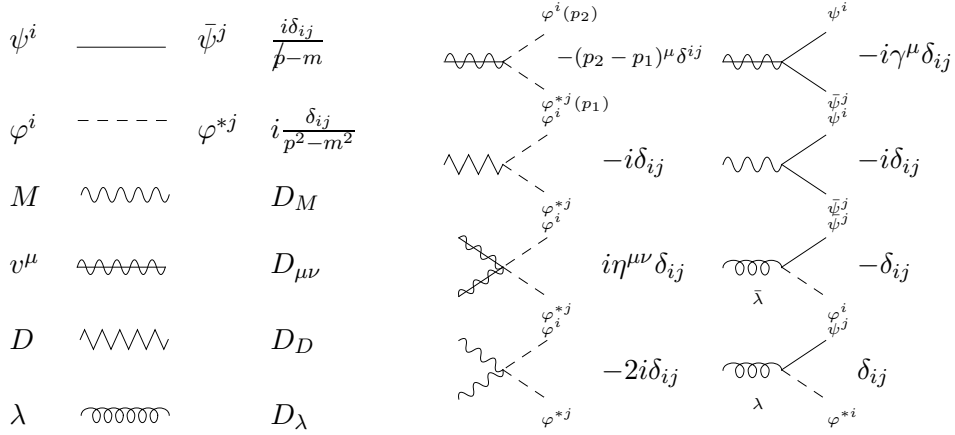
\begin{figure}[h]
\begin{center}
\unitlength=1mm
\begin{picture}(110,60)

\includegraphics[width=10cm]{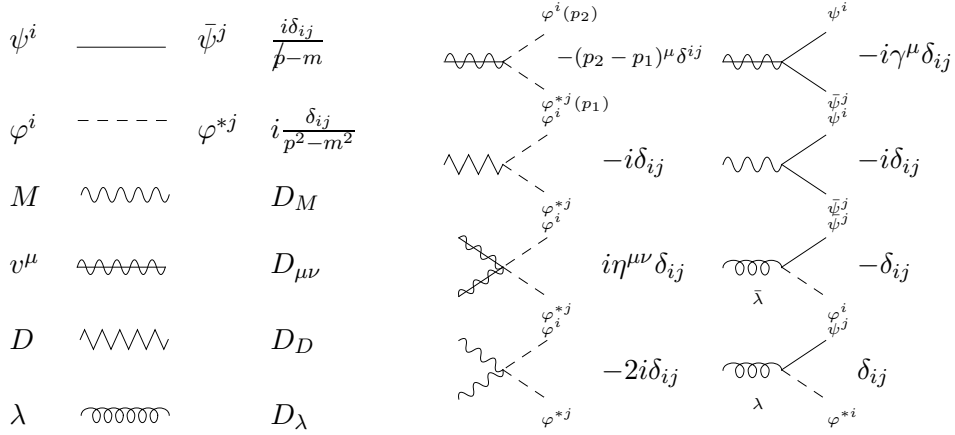}
\put(-110,50){ $\psi^i$}
\put(-85,50){ $\bar{\psi}^j$}
\put(-74,50){$\frac{i\delta_{ij}}{p\llap /-m}$}

\put(-110,38){ $\varphi^i$}
\put(-85,38){ $\varphi^{*j}$}
\put(-74,38){$i \frac{\delta_{ij}}{p^2-m^2}$}

\put(-110,29){ $M$}
\put(-74,29){$D_M$}

\put(-110,20){ $v^\mu$}
\put(-74,20){$D_{\mu \nu}$}

\put(-110,10){ $D$}
\put(-74,10){$D_{D}$}

\put(-110,0){ $\lambda$}
\put(-74,0){$D_{\lambda}$}

\put(-36,48){\scriptsize$-(p_2-p_1)^\mu \delta^{ij}$}
\put(-38,54){\tiny$\varphi^i (p_2)$}
\put(-38,42){\tiny$\varphi^{* j} (p_1)$}

\put(-30,34){$-i\delta_{ij}$}
\put(-38,40){\tiny$\varphi^i $}
\put(-38,28){\tiny$\varphi^{* j} $}

\put(-30,20){$i \eta^{\mu \nu} \delta_{ij}$}
\put(-38,26){\tiny$\varphi^i $}
\put(-38,14){\tiny$\varphi^{* j} $}

\put(-30,6){$-2 i \delta_{ij}$}
\put(-38,12){\tiny$\varphi^i $}
\put(-38,0){\tiny$\varphi^{* j} $}

\put(4,48){$-i \gamma^\mu \delta_{ij}$}
\put(0,54){\tiny$\psi^i $}
\put(0,42){\tiny$\bar{\psi}^{ j} $}

\put(4,34){$- i \delta_{ij}$}
\put(0,40){\tiny$\psi^i $}
\put(0,28){\tiny$\bar{\psi}^{ j} $}

\put(4,20){$-\delta_{ij}$}
\put(-10,16){\tiny$\bar{\lambda}$}
\put(0,26){\tiny$\bar{\psi}^j $}
\put(0,14){\tiny$\varphi^{ i} $}

\put(4,6){$\delta_{ij}$}
\put(-10,2){\tiny$\lambda$}
\put(0,12){\tiny$\psi^{j} $}
\put(0,0){\tiny$\varphi^{ *i} $}

\end{picture}
\caption{The Feynman rules of $CP^{N-1}$ model in Minkowski spaces}

\end{center}
\end{figure}
Here,
\beq
D_M&=&\frac{ 4\pi i}{N}\frac{1}{p^2-4m^2} I(p^2),\nonumber\\
D_{\mu \nu}&=&\frac{ 4\pi}{N}\frac{1}{p^2-4m^2}[-i \eta_{\mu \nu} +(1+\alpha+\frac{4m^2 \alpha}{p^2})\frac{i p_\mu p_\nu}{p^2} +\frac{2m \epsilon_{\mu \nu \lambda} p^\lambda}{p^2}] I(p^2),\nonumber\\
D_D&=&\frac{ 4\pi i}{N} I(p^2),\nonumber\\
D_{\lambda}&=&-\frac{ 8\pi i}{N} \frac{1}{p\llap /+2m} I(p^2),\nonumber\\
I(p)&=&\frac{\sqrt{-p^2}}{\arctan(\sqrt{\frac{-p^2}{4m}})}.\nonumber
\eeq
\newpage

\subsection{The $Q^{N-2}$ model}\label{feynman-QN}
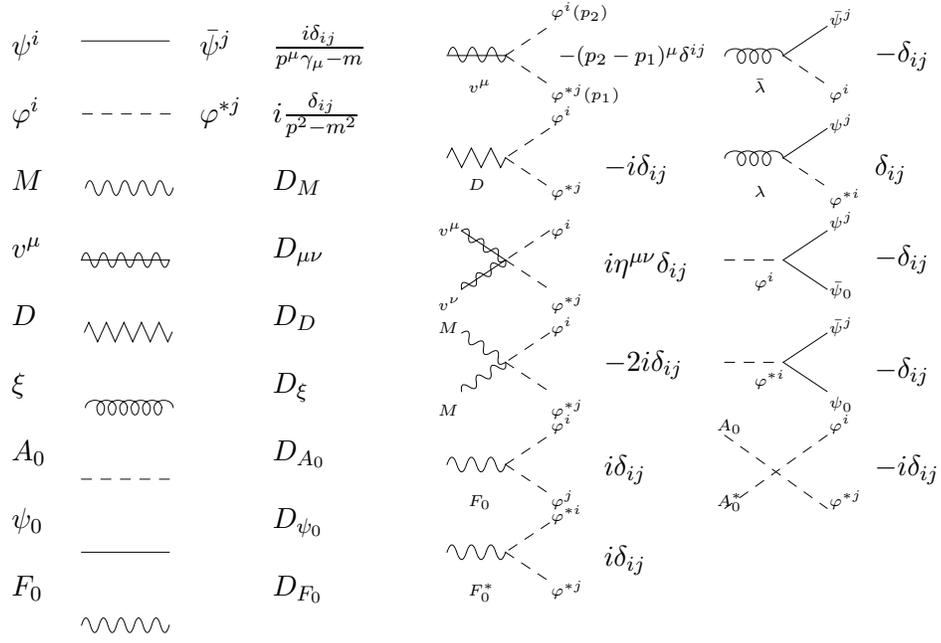
\begin{figure}[h]
\begin{center}
\unitlength=1mm
\begin{picture}(90,90)

\includegraphics[width=10cm]{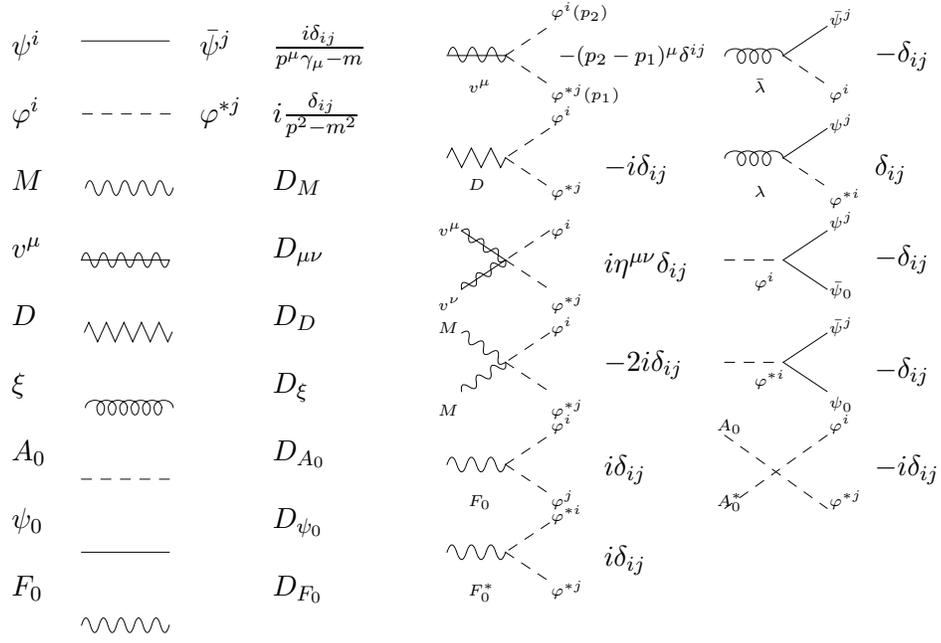}
\put(-110,78){ $\psi^i$}
\put(-85,78){ $\bar{\psi}^j$}
\put(-74,78){$\frac{i\delta_{ij}}{p^\mu \gamma_\mu-m}$}

\put(-110,69){ $\varphi^i$}
\put(-85,69){ $\varphi^{*j}$}
\put(-74,69){$i \frac{\delta_{ij}}{p^2-m^2}$}

\put(-110,60){ $M$}
\put(-74,60){$D_M$}

\put(-110,51){ $v^\mu$}
\put(-74,51){$D_{\mu \nu}$}

\put(-110,42){ $D$}
\put(-74,42){$D_{D}$}

\put(-110,33){ $\xi$}
\put(-74,33){$D_{\xi}$}

\put(-110,24){ $A_0$}
\put(-74,24){$D_{A_0}$}

\put(-110,15){ $\psi_0$}
\put(-74,15){$D_{\psi_0}$}

\put(-110,6){ $F_0$}
\put(-74,6){$D_{F_0}$}

\put(-36,77){\scriptsize$-(p_2-p_1)^\mu \delta^{ij}$}
\put(-48,73){\tiny$v^\mu$}
\put(-37,83){\tiny$\varphi^i (p_2)$}
\put(-37,72){\tiny$\varphi^{* j} (p_1)$}

\put(-30,62){$-i\delta_{ij}$}
\put(-48,60){\tiny$D$}
\put(-37,69){\tiny$\varphi^i $}
\put(-37,59){\tiny$\varphi^{* j} $}

\put(-30,49){$i \eta^{\mu \nu} \delta_{ij}$}
\put(-52,54){\tiny$v^\mu$}
\put(-52,44){\tiny$v^\nu$}
\put(-37,54){\tiny$\varphi^i $}
\put(-37,44){\tiny$\varphi^{* j} $}

\put(-30,36){$-2 i \delta_{ij}$}
\put(-52,41){\tiny$M$}
\put(-52,30){\tiny$M$}
\put(-37,41){\tiny$\varphi^i $}
\put(-37,30){\tiny$\varphi^{* j} $}

\put(-30,22){$i\delta_{ij}$}
\put(-48,18){\tiny$F_0$}
\put(-37,28){\tiny$\varphi^i $}
\put(-37,18){\tiny$\varphi^{ j} $}

\put(-30,10){$i\delta_{ij}$}
\put(-48,6){\tiny$F_0^*$}
\put(-37,16){\tiny$\varphi^{*i} $}
\put(-37,6){\tiny$\varphi^{* j} $}


\put(6,77){$-\delta_{ij}$}
\put(-10,73){\tiny$\bar{\lambda}$}
\put(0,82){\tiny$\bar{\psi}^j $}
\put(0,72){\tiny$\varphi^{ i} $}

\put(6,62){$\delta_{ij}$}
\put(-10,59){\tiny$\lambda$}
\put(0,68){\tiny$\psi^{j} $}
\put(0,58){\tiny$\varphi^{ *i} $}

\put(6,50){$-\delta_{ij}$}
\put(-10,47){\tiny$\varphi^i$}
\put(0,55){\tiny$\psi^{j} $}
\put(0,46){\tiny$\bar{\psi}_0 $}

\put(6,35){$-\delta_{ij}$}
\put(-10,34){\tiny$\varphi^{*i}$}
\put(0,41){\tiny$\bar{\psi}^{j} $}
\put(0,31){\tiny$\psi_0 $}

\put(6,22){$- i \delta_{ij}$}
\put(-15,28){\tiny$A_0$}
\put(-15,18){\tiny$A_0^*$}
\put(0,28){\tiny$\varphi^i $}
\put(0,18){\tiny$\varphi^{* j} $}

\end{picture}
\caption{The Feynman rule of $Q^{N-2}$ model in Minkowski spaces. We show only the parts which are concerned the next-to-leading correction of $\varphi^i$ propagator.}

\end{center}
\end{figure}
Here,
\beq
D_M&=&\frac{ 4\pi i}{N}\frac{1}{p^2-4m^2} I(p^2),\nonumber\\
D_{\mu \nu}&=&\frac{ 4\pi}{N}\frac{1}{p^2-4m^2}[-i \eta_{\mu \nu} +(1+\alpha+\frac{4m^2 \alpha}{p^2})\frac{i p_\mu p_\nu}{p^2} +\frac{2m \epsilon_{\mu \nu \lambda} p^\lambda}{p^2}] I(p^2),\nonumber\\
D_D&=&\frac{ 4\pi i}{N} I(p^2),\nonumber\\
D_{\xi}&=&-\frac{ 8\pi i}{N} \frac{1}{p^\mu \gamma_\mu+2m} I(p^2),\nonumber\\
D_{A_0}&=&\frac{8 \pi i}{N}\frac{1}{p^2-4 m^2} I(p^2),\nonumber\\
D_{\psi_0}&=&\frac{8 \pi i}{N} \frac{1}{p^\mu \gamma_\mu +2m} I(p^2)\nonumber\\
D_{F_0}&=&\frac{8 \pi i}{N}I(p^2).\nonumber
\eeq

\section{The broken phase of $Q^{N-2}$ model} \label{QN-broken}
From the stationaly condition of the effective potential, we obtain the set of vacuum expectation values,
\beq
M_0=0, \hspace{0.1cm}|A_0|=0, \hspace{0.1cm}|D_1|^{\frac{1}{2}}=\frac{4 \pi}{N}\varphi_1^2+m, \hspace{0.1cm}|D_2|^{\frac{1}{2}}=\frac{4 \pi}{N}\varphi_2^2 +m.
\eeq
Then the effective potential becomes
\beq
V_{eff}=\frac{1}{24 \pi} \Big[ |\frac{4\pi}{N} \varphi_1^2 +m|^3 +|\frac{4 \pi}{N}\varphi_2^2 +m|^3  \Big] ,
\eeq
and the minimal value of this effective potential is realized when 
\beq
\varphi_1^2=\frac{N}{4 \pi}|m|, \hspace{0.2cm} \varphi_2^2=\frac{N}{4 \pi}|m|.
\eeq
Using $SO(N)$ symmetry and the constraint $\varphi_1 \varphi_2=0$ which is obtained by the stationary condition, we choose the vacuum expectation values as follow:
\begin{align}
\varphi_i = \begin{pmatrix}
         0  \\
        \vdots \\
         0  \\
        i \sqrt{\frac{N}{4 \pi}|m|} \\
        \sqrt{\frac{N}{4 \pi}|m|} \\
        \end{pmatrix}.
\end{align}
Thus global $SO(N)$ symmetry breaks to $SO(N-2)$.

Futhermore, we put the fluctuation rewritten the $\varphi^{N-1}$ and $\varphi^{N}$, and these superpartnars $\psi^{N-1}$ and $\psi^{N}$ as
\beq
\frac{1}{\sqrt{2}} (\varphi^N -i \varphi^{N-1})={\cal A}_N,\nonumber\\
\frac{1}{\sqrt{2}} (\varphi^N +i \varphi^{N-1})={\cal A}_{N-1},\nonumber\\
\frac{1}{\sqrt{2}} (\psi^N -i \psi^{N-1})=\Psi_N,\nonumber\\
\frac{1}{\sqrt{2}} (\psi^N +i \psi^{N-1})=\Psi_N.\nonumber
\eeq
Then the effective action expanding up to the next-to-leading order of $1/N$ can be written by
\begin{align}
S_{eff}&=
\int \! \frac{d^3 p}{(2 \pi)^3}
 \biggl[ \Bigl\{
  A_0^*(-p) ( - \frac{N}{16} \sqrt{- p^2} - \frac{N}{4 \pi}|m| )  A_0(p) 
+ \frac{1}{2} M(-p) ( - \frac{N}{8}  \sqrt{- p^2} - \frac{N}{2 \pi}|m| )  M(p)
\nonumber \\
&- \frac{1}{2} v^{\mu}(-p) \left( \frac{N}{8} \frac{p^2 \eta_{\mu \nu} - p_{\mu} p_{\nu} }
{\sqrt{-p^2}} - \frac{1}{2 \pi}|m| \right) 
 v^{\nu}(p)  
 + \bar{\lambda}(-p) \frac{N}{16} \frac{p^\mu \gamma_\mu}{\sqrt{-p^2}} 
 \lambda(p)
\nonumber \\
&+ \bar{\psi}_0(-p) \frac{N}{16} \frac{p^\mu \gamma_\mu}{\sqrt{-p^2}}  \psi_0(p)  
+ \frac{1}{2} D(-p)\frac{N}{8 \sqrt{-p^2}}  D(p) 
+ F_0^*(-p) \frac{N}{16 \sqrt{-p^2}}  F_0(p) \Bigr\}
 \nonumber \\ 
\ & \ + \sum_{k=1}^{N-2} A_k^{*}(-p) p^2 A_k(p) 
+ {\cal A}_{N-1}^*(-p) p^2 {\cal A}_{N-1}(p)
+ {\cal A}_N^*(-p) p^2 {\cal A}_N(p)\nonumber\\
 & + \sqrt{\frac{N}{4 \pi}|m|} 
\left( D(-p) {\cal A}_N(p) + D(-p) {\cal A}_N^*(p)  
 \right)  + \sqrt{\frac{N}{4 \pi}|m|} 
\left\{  F_0(-p) {\cal A}_{N-1}(p) + F_0^*(-p) {\cal A}_{N-1}^*(p)
\right\} \nonumber \\
&+ \sum_{k=1}^{N-2} \bar{\psi}_k(-p) p^\mu \gamma_\mu \psi_k(p)
+ \bar{\Psi}_{N-1}(-p) p^\mu \gamma_\mu \Psi_{N-1}(p)
+ \bar{\Psi}_{N}(-p) p^\mu \gamma_\mu \Psi_{N}(p)\nonumber\\
& - i  \sqrt{\frac{N}{4 \pi}|m|} 
 \left(  \bar{\Psi}_N(-p) \bar{\lambda}(p)
 - \Psi_{N}(-p) \lambda (p)
 \right)
  - \sqrt{\frac{N}{4 \pi}|m|} \left( 
           \bar{\psi}_0(-p) \bar{\Psi}_{N-1}(p) 
          + \psi_0(-p) \Psi_{N-1}(p) 
\right)
\nonumber \\
\ & \  - \sqrt{\frac{N}{4 \pi}|m|} v^{\mu}(-p) p_{\mu} \left( 
          {\cal A}_N(p)  - {\cal A}_N^*(p) 
\right)
 +({\rm interactions})  \biggr] \; .
\end{align}
This effective action is similar to the one of $CP^{N-1}$ broken phase.
We put ${\cal A}_N=\frac{1}{\sqrt{2}}({\cal A}_N^R +i {\cal A}_N^I)$ and rewrite the effective action as 
\begin{gather}
S_{eff}= \int \frac{d^3 p}{(2 \pi)^3} \frac{1}{2} ( D \  {\cal A}_N^R ) 
\begin{pmatrix}
\frac{N}{8 \sqrt{-p^2}} & \sqrt{\frac{N}{2 \pi}|m|} \\
\sqrt{\frac{N}{2 \pi}|m|} & p^2
\end{pmatrix}
\begin{pmatrix}
D \\
{\cal A}_N^R
\end{pmatrix} \;  \\
+( F_0 \  {\cal A}_{N-1}^* ) 
\begin{pmatrix}
\frac{N}{16 \sqrt{-p^2}} & \sqrt{\frac{N}{4 \pi}|m|} \\
\sqrt{\frac{N}{4 \pi}|m|} & p^2
\end{pmatrix}
\begin{pmatrix}
F_0 \\
{\cal A}_{N-1}
\end{pmatrix} \\
+( \bar{\lambda} \  \Psi_N ) 
\begin{pmatrix}
\frac{N p^\mu \gamma_\mu}{16 \sqrt{-p^2}} & -i \sqrt{\frac{N}{4 \pi}|m|} \\
i \sqrt{\frac{N}{4 \pi}|m|} & p^\mu \gamma_\mu
\end{pmatrix}
\begin{pmatrix}
\lambda \\
\bar{\Psi}_N
\end{pmatrix} \\
+( \bar{\psi}_0 \  \Psi_{N-1} ) 
\begin{pmatrix}
\frac{N p^\mu \gamma_\mu}{16 \sqrt{-p^2}} & - \sqrt{\frac{N}{4 \pi}|m|} \\
- \sqrt{\frac{N}{4 \pi}|m|} & p^\mu \gamma_\mu
\end{pmatrix}
\begin{pmatrix}
\psi_0 \\
\bar{\Psi}_{N-1}
\end{pmatrix}
 \cdots.
\end{gather}
The eigen value of these matrix has no solution, then we find these fields are not physical in this phase.


\end{document}